\documentclass[amsmath,amssymb,amsfonts,aps,pre,preprint,superscriptaddress,bibnotes,showpacs,showkeys,longbibliography,nofootinbib]{revtex4-1}
\usepackage{mathtools}
\usepackage{epsfig}
\usepackage[english]{babel}
\usepackage{physics}
\usepackage{color}
\usepackage{subcaption}
\usepackage{hyperref}
\usepackage{lipsum}
\usepackage{cases}

\AtBeginDocument{%
    \newwrite\bibnotes
    \def\bibnotesext{Notes.bib}
    \immediate\openout\bibnotes=\jobname\bibnotesext
    \immediate\write\bibnotes{@CONTROL{REVTEX41Control}}
    \immediate\write\bibnotes{@CONTROL{%
    apsrev41Control,author="08",editor="1",pages="1",title="0",year="1"}}
     \if@filesw
     \immediate\write\@auxout{\string\citation{apsrev41Control}}%
    \fi
}%

\begin{document}
\title{Coastlines and percolation in a model for hierarchical random deposition}

\author{Jonas Berx}
\affiliation{Institute for  Theoretical Physics, KU Leuven, B-3001 Leuven, Belgium}

\author{Evi Bervoets}
\affiliation{Institute for  Theoretical Physics, KU Leuven, B-3001 Leuven, Belgium}

\author{Claudiu V. Giuraniuc}
\affiliation{Institute of Medical Sciences, University of Aberdeen, Aberdeen, Scotland, UK}

\author{Joseph O. Indekeu\footnote{Given his role as Editor of this journal, Joseph O. Indekeu had no involvement in the peer-review of articles for which he was an author and had no access to information regarding their peer-review. Full responsibility for the peer-review process for this article was delegated to another Editor.}}
\affiliation{Institute for  Theoretical Physics, KU Leuven, B-3001 Leuven, Belgium}

\date{\today}

\begin{abstract}
We revisit a known model in which (conducting) blocks are hierarchically and randomly deposited on a $d$-dimensional substrate according to a hyperbolic size law with the block size decreasing by a factor $\lambda \, >  1$ in each subsequent generation. In the first part of the paper the number of coastal points (in $D=1)$ or coastlines (in $D=2)$ is calculated, which are points or lines that separate a region at ``sea level" and an elevated region. We find that this number possesses a non-universal character, implying a Euclidean geometry below a threshold value $P_c$ of the deposition probability $P$, and a fractal geometry above this value. Exactly at the threshold, the geometry is logarithmic fractal. The number of coastlines in $D=2$ turns out to be exactly twice the number of coastal points in $D=1$. We comment briefly on the surface morphology and derive a roughness exponent $\alpha$. In the second part, we study the percolation probability for a current in this model and two extensions of it, in which both the scale factor and the deposition probability can take on different values between generations. We find that the percolation threshold $P_c$ is located at exactly the same value for the deposition probability as the threshold probability of the number of coastal points. This coincidence suggests that exactly at the onset of percolation for a conducting path, the number of coastal points exhibits logarithmic fractal behaviour. 

\end{abstract}

\maketitle

\section{Introduction}

While both Euclidean and fractal geometry  are well-known to physicists, the borderline case where the Hausdorff-Besicovitch dimension $D_f$ equals the topological (or Euclidean) dimension $D$ has received much less attention in the literature. A subset of these cases can be characterised by the fractal measure with ruler length $\rho$
\begin{equation}
    \label{eq:measure}
    h(\rho) = \rho^{D_f} \left[\log{(1/\rho)}\right]^{\Delta_1}\, , \; \mbox{with} \; D_f = D,
\end{equation}
where $\Delta_1$ is a subdimension \cite{Mandelbrot}. The occurrence of the logarithm in the fractal measure has led  researchers \cite{INDEKEU1998294} to coin the term \textsl{logarithmic fractals}. These objects can be distinguished from classical fractals by observing that quantities such as length and area increase linearly instead of exponentially, upon decreasing the ``ruler length". The same research has shown that the logarithmic fractal behaviour of a basic model for hierarchical deposition is robust under randomness and that the surface length or area, asymptotically for large generation number, increases by a constant. 

One can study level sets for random hierarchical deposition on a Euclidean substrate of dimension $D$ and the resulting collection of points or contour lines, which are, respectively, named coastal points and coastlines for  $D=1$ or $D=2$ \cite{INDEKEU2000135}. This nomenclature stems from the context of islands, which result when the landscape is floaded up to a certain level, and is useful for describing the geometry of their jagged beaches. Real-world applications of fractal level sets come to mind when considering such geometries, some examples include the flooding of Arctic melt ponds \cite{Bowen2018} or the fractal growth of thin metallic films \cite{AMINIRASTABI2020122261} or of bacterial populations \cite{INDEKEUSZNAJD,INDEKEU200414}.

The study of level sets is deeply connected to the theory of percolation \cite{stauffer_1992,STAUFFER19791}, in which the geometry of the system under consideration changes drastically when the deposition probability reaches a critical value named the \textsl{percolation threshold}. When percolation is achieved, one expects the behaviour of the number of coastal points/coastlines to change. We will show that in the random hierarchical deposition model the percolation threshold indicates a transition from a Euclidean to a fractal geometry. Exactly at the percolation threshold, however, the number of coastal points exhibits a logarithmic fractal behaviour, growing linearly with increasing generation number.

\section{Hierarchical deposition}\label{sec:1D}

In this section we recapitulate briefly the setup and some of the elementary properties of the hierarchical deposition model in  $D=1$ (and $D=2$). Let us consider the deposition of squares (or the digging of holes) on a line $[0,1]$  which fall down in a temporal order determined by their size. The largest ones fall down first, followed by squares of which the sides are smaller by a factor $\lambda$. We assume a hyperbolic distribution of the number of squares deposited according to their size. The number of squares of linear size $s$ is denoted by $N(s)$ and obeys the following scaling,
\begin{equation}
    \label{eq:square_distribution}
    N(s) = \lambda^{-1} N(s/\lambda)\, 
\end{equation}
where $\lambda\in\mathbb{R}^+$ and $\lambda > 1$. One can see that the number of squares is proportional to the inverse impact cross-section. In this model \cite{INDEKEU1998294,INDEKEU2000135} a logarithmic fractal law was found for the surface (length) of the resulting landscape.

Let us first recapitulate the above model for \textsl{random} deposition, as was investigated in \cite{INDEKEU1998294}. First, divide the unit interval in $\lambda$ subsets of length $1/\lambda$, where now $\lambda\in\mathbb{N}$ and either deposit a square ``hill" in each of the subsets with probability $P$ or dig a square ``hole" with probability $Q$, which reduces the height with a factor $\lambda$. The third option is to do nothing with probability $1-S \equiv 1-(P+Q)$. In the second generation ($n=2)$ we divide each of the subsets of the previous generation in smaller subsets with length $1/\lambda^2$ and start depositing blocks or digging holes once again. This process can be repeated indefinitely and the resulting asymptotic fractal properties for the generation number $n\rightarrow\infty$ can be studied. 

It has been shown in \cite{INDEKEU1998294} that for $\lambda \geq 3$, the asymptotic surface length increment $\Delta L_\infty$ for infinite generation number $n$ has the following form
\begin{equation}
    \label{eq:length_increment}
    \Delta L_\infty = \frac{2 \left[P(1-P)+Q(1-Q)\right]}{1+2 \left[P(1-P)+Q(1-Q)-PQ\right]/(\lambda-1)}\, ,
\end{equation}
while for $\lambda=2$, partial levelling of vertical segments of the landscape increases the resulting surface length increment somewhat \cite{INDEKEU1998294}.

These results are easily extended to one substrate dimension higher \cite{INDEKEU1998294}. The main difference with respect to the situation for $D = 1$ is that in $D=2$ the number of cubes of linear size $s$ is inversely proportional to their cross-sectional area $s^2$, i.e.,
\begin{equation}
    \label{eq:square_distribution_cubes}
    N(s) = \lambda^{-2} N(s/\lambda)\, 
\end{equation}
and that in generation $n$ a wall that was put in generation $m$ occupies $\lambda^{n-m}$ edges. The surface is now divided into square \textsl{plaquettes}. Neighbouring plaquettes share an edge. In generation $n$ there are $\lambda^{2n}$ plaquettes and twice as many edges. After a careful inspection and calculation one concludes that the (dimensionless) area increment for $D=2$ obeys a law similar to that which is satisfied by the (dimensionless) length increment for $D=1$. The substrate directions manifest themselves as independent. Consequently, in dimensionless units of reduced  area, ons obtains twice the result of equation \eqref{eq:length_increment} for $D=2$, i.e., the \textsl{area} increment is related to the \textsl{length} increment through \cite{INDEKEU1998294},
\begin{equation}
    \label{eq:surface_area_increment}
    \Delta S_\infty = 2 \Delta L_\infty \, .
\end{equation}

\section{Coastal points and non-universality} \label{sec:coastal}
A richer variety of phenomena appears when level sets of the deposition model are studied \cite{INDEKEU2000135}. We define a ``sea level" in the landscape and study the geometry of the coastal points in $D=1$ (or coastlines in $D=2$) that remain. If the substrate upon which the squares are placed is a line, the level set at sea level is the number of coastal points. For $D=2$, i.e., the substrate is planar, the level set is the total length of the coastlines of the resulting islands. In the following discussion, we choose the sea level to be close to the zero level of the substrate. We will present here the case for the one-dimensional model in which the substrate is a line, generalization to higher dimensions is straightforward.

\begin{figure}[ht]
    \centering
    \includegraphics[width = 0.6\linewidth]{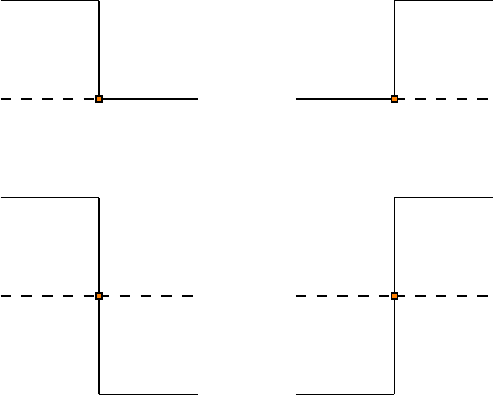}
    \caption{Definition of coastal points. The dotted line indicates sea level, while full lines represent either a hill or a hole. The coastal points are drawn as orange squares.}
    \label{fig:coastal_points_def}
\end{figure}

Consider the unit interval $[0,1]$ with rescaling factor $\lambda\geq2$ and probabilities $P,Q$, respectively, to deposit a hill or dig a hole. In generation $n$, we will call a \textsl{coastal point} a meeting point at sea level between two segments of length $\lambda^{-n}$, one of which supports a newly placed hill above sea level and the other marking a newly dug hole below sea level or being a segment that has remained at sea level, and vice versa, as shown in Fig. \ref{fig:coastal_points_def}. Points that are created and are not coastal points we call \textsl{internal points}. Points that remain from a previous generation without being coastal or internal points are called \textsl{external points}. 

\begin{figure}[h]
        \centering
            \begin{subfigure}[b]{0.4\textwidth}
                 \centering
                 \includegraphics[width=\textwidth]{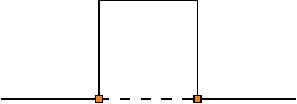}
                 \label{fig:CPA1}
            \end{subfigure}\hspace{2cm}
            \begin{subfigure}[b]{0.4\textwidth}
                 \centering
                 \includegraphics[width=\textwidth]{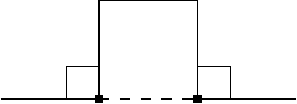}
                 \label{fig:CPA2}
            \end{subfigure}
            \caption{Destruction of coastal points (orange squares) in generation $n-1$ (left) by deposition of a smaller hill on sea level, lifting the point in generation $n$ (right).}
            \label{fig:CPA}
    \end{figure}

New coastal points can be created in generation $n$ in a number of different ways \cite{MscBervoets,PhdGiuraniuc}:
\begin{enumerate}
    \item Internal points are formed in generation $n$ on a segment which was at sea level in generation $n-1$, see Fig. \ref{fig:CP12}. 
    \begin{figure}
        \centering
            \begin{subfigure}[b]{0.4\textwidth}
                 \centering
                 \includegraphics[width=\textwidth]{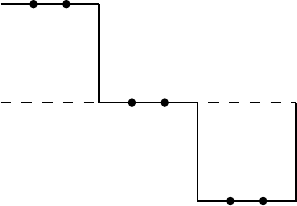}
                 \label{fig:CP1}
            \end{subfigure}\hspace{2cm}
            \begin{subfigure}[b]{0.4\textwidth}
                 \centering
                 \includegraphics[width=\textwidth]{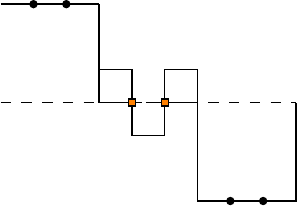}
                 \label{fig:CP2}
            \end{subfigure}
            \caption{Creation of coastal points (orange squares) in generation $n$ (right) by deposition on internal points (black circles) that were still at sea level in generation $n-1$ (left). }
            \label{fig:CP12}
    \end{figure}
    These internal points from all previous generations generate on average the following number of coastal points
    \begin{equation}
        \label{eq:coastal_A}
        \lambda(\lambda-1)(1-S) \sum\limits_{i=2}^n \left[\lambda (1-S)\right]^{i-2} 2P (1-P)\, .
    \end{equation}
    \item External points at sea level that did not experience any deposition up until generation $n-1$, see Fig. \ref{fig:CP34}. 
    \begin{figure}[h]
        \centering
            \begin{subfigure}[b]{0.4\textwidth}
                 \centering
                 \includegraphics[width=\textwidth]{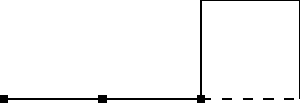}
                 \label{fig:CP3}
            \end{subfigure}\hspace{2cm}
            \begin{subfigure}[b]{0.4\textwidth}
                 \centering
                 \includegraphics[width=\textwidth]{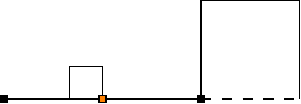}
                 \label{fig:CP4}
            \end{subfigure}
            \caption{Creation of coastal points (orange square) in generation $n$ (right) by deposition on external points (black squares) that were still at sea level in generation $n-1$ (left). }
            \label{fig:CP34}
    \end{figure}
    
    These external points generate on average \begin{equation}
        \label{eq:coastal_B}
        \lambda \sum\limits_{i=1}^n \left[(1-S)^2\right]^{i-1} 2P(1-P) + \lambda (\lambda-1)(1-S)^3 \left(\sum\limits_{k=3}^n \sum\limits_{i=3}^k \lambda^{k-i} (1-S)^{k-6+i}\right)2P(1-P)
    \end{equation}
    coastal points in generation $n$. 
    \item External points at sea level which at generation $n-1$ connect a segment at sea level with a hole. Placing a hill on sea level next to the hole in generation $n$ creates the coastal point, see Fig. \ref{fig:CP56}.
    \begin{figure}[h]
        \centering
            \begin{subfigure}[b]{0.4\textwidth}
                 \centering
                 \includegraphics[width=\textwidth]{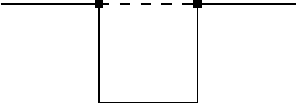}
                 \label{fig:CP5}
            \end{subfigure}\hspace{2cm}
            \begin{subfigure}[b]{0.4\textwidth}
                 \centering
                 \includegraphics[width=\textwidth]{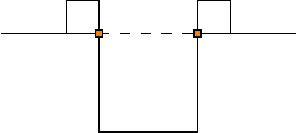}
                 \label{fig:CP6}
            \end{subfigure}
            \caption{Creation of coastal points (orange squares) in generation $n$ (right) by deposition on external points (black squares) which at generation $n-1$ (left) connect a segment at sea level with a hole. }
            \label{fig:CP56}
    \end{figure}
    This procedure yields the following number of coastal points
    \begin{equation}
        \label{eq:coastal_C}
         2 \lambda Q (1-S)\left(\sum\limits_{k=2}^n \sum\limits_{i=2}^k \lambda^{k-i} (1-S)^{k-4+i}\right)P\, .
    \end{equation}
\end{enumerate}
Coastal points can also be destroyed in generation $n$ by placing a smaller hill on a sea level segment right next to an existing hill, hereby lifting the point. This is illustrated in Fig. \ref{fig:CPA}.

On average, the number of coastal points that get destroyed in generation $n$ is equal to 
\begin{equation}
        \label{eq:coastal_D}
         2 \lambda P (1-S)\left(\sum\limits_{k=2}^n \sum\limits_{i=2}^k \lambda^{k-i} (1-S)^{k-4+i}\right)P\, .
\end{equation}
Finally, adding equations \eqref{eq:coastal_A}, \eqref{eq:coastal_B}, \eqref{eq:coastal_C} and subtracting equation \eqref{eq:coastal_D} results in a complete expression for the number of coastal points $\mathcal{N}_n(P,Q)$ in generation $n$ as a function of the deposition probabilities $P$ and $Q$,
\begin{equation}
    \label{eq:coastal_points}
    \begin{split}
        \mathcal{N}_n(P,Q) &= \lambda \sum\limits_{i=1}^n \left[(1-S)^2\right]^{i-1} 2P(1-P)\\
        &+ \lambda(\lambda-1)(1-S) \sum\limits_{i=2}^n \left[\lambda (1-S)\right]^{i-2} 2P (1-P)\\
        &+ 2 \lambda (Q-P) (1-S)\left(\sum\limits_{k=2}^n \sum\limits_{i=2}^k \lambda^{k-i} (1-S)^{k-4+i}\right)P\\
        &+ \lambda (\lambda-1)(1-S)^3 \left(\sum\limits_{k=3}^n \sum\limits_{i=3}^k \lambda^{k-i} (1-S)^{k-6+i}\right)2P(1-P)\, .
    \end{split}
\end{equation}

\begin{figure}[htp]
    \centering
    \includegraphics[width=0.7\linewidth]{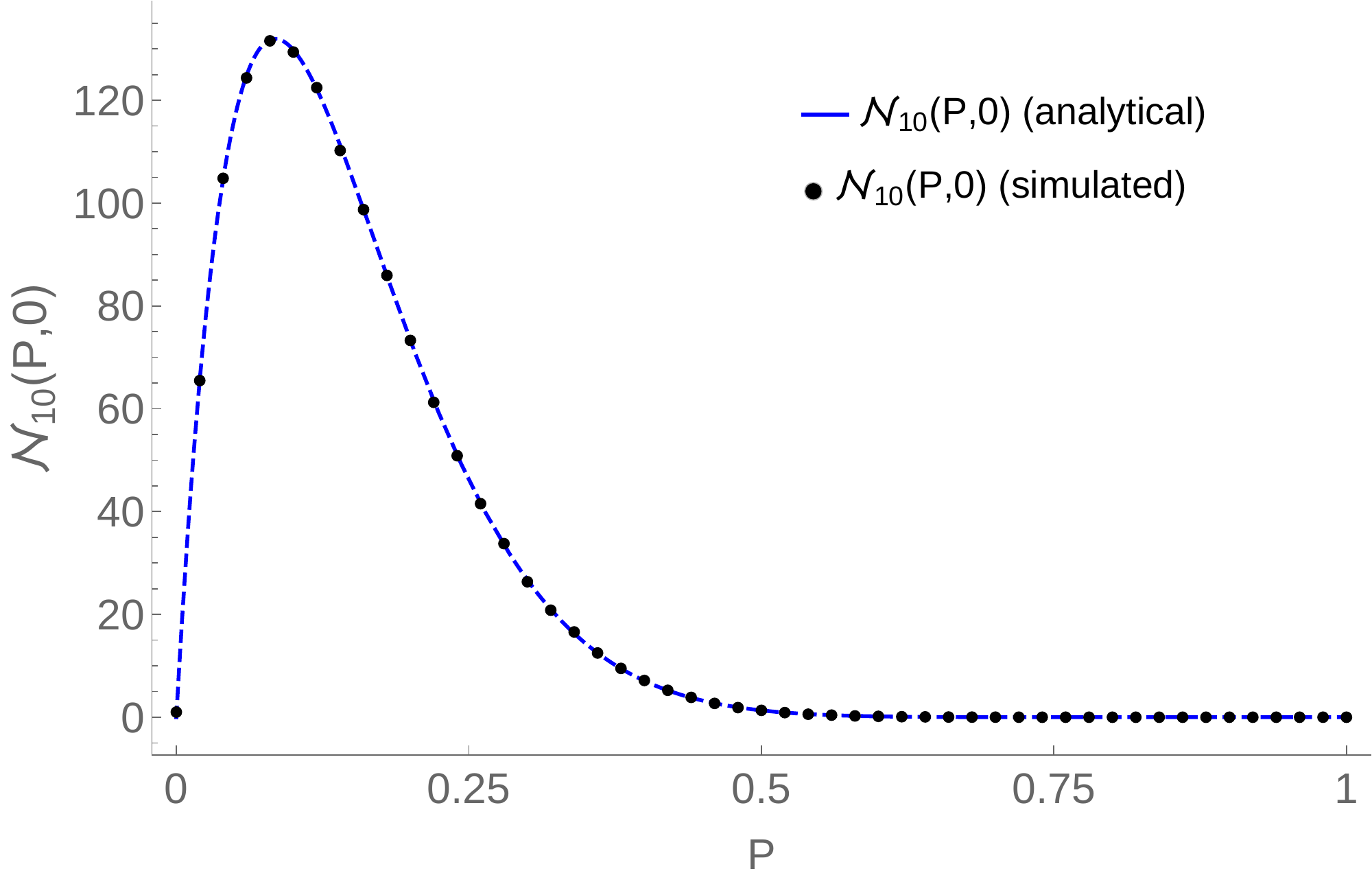}
    \caption{The number of coastal points $\mathcal{N}_{10}(P,0)$ in generation $n=10$ for $\lambda=2$ and $Q=0$. The numerical results are shown (black dots) together with the theoretical prediction \eqref{eq:coastal_points} (blue, dashed line). The numerical results were averaged over 20000 runs.}
    \label{fig:coastal_points_solution}
\end{figure}

The number of coastal points \eqref{eq:coastal_points} for generation $n=10$ are shown in Fig. \ref{fig:coastal_points_solution} for $\lambda = 2$ and $Q = 0$ together with numerical results from simulations which were performed using direct simulation with a random number generator. The sums in expression \eqref{eq:coastal_points} can be worked out exactly and the expression can be simplified.  The details of these calculations can be found in the appendix \ref{app:1}. From these calculations one can distinguish three regimes for the number of coastal points:
\begin{itemize}
    \item Euclidean regime for $\lambda(1-S)<1$. Here the number of coastal points saturates to a constant value $\gamma$, see equation \eqref{eq:gamma}. 
    \item Fractal regime with nontrivial fractal dimension $0<D_f<1$ for $\lambda(1-S)>1$. In this regime the number of coastal points increases exponentially. 
    \item Logarithmic fractal regime at $\lambda(1-S)=1$. The number of coastal points increases linearly with increasing generation $n$.
\end{itemize}
This non-universality of the coastal points is in sharp contrast with the universal fractal properties of the surfaces described in Section \ref{sec:1D}, which are always marginally fractal, independent of the model parameters. In two dimensions ($D=2$), the previously obtained expression for the number of coastal points \eqref{eq:coastal_points} is to be multiplied by a factor 2 in order to obtain the (dimensionless) length of the coastline.

The fractal dimension $D_f$ can be calculated by considering the increase of the number of coastal points with  increasing generation number $n$, i.e.,
\begin{equation}
    \label{eq:fractal_dimension}
    D_f = \lim_{n\rightarrow\infty}\frac{\ln{\mathcal{N}_n(P,Q)}}{\ln{\lambda^n}} = \frac{\ln{\left[\lambda (1-S)\right]}}{\ln{\lambda}} =  1+\frac{\ln{(1-S)}}{\ln{\lambda}}
\end{equation}
for deposition on a line. When the substrate is planar ($D=2$), the above expression \eqref{eq:fractal_dimension} is augmented by 1. This gives, respectively, $0<D_f<1$ and $1<D_f<2$ for deposition on a line or on a plane.

\section{The critical exponents of the resulting surface}
\label{sec:exponents}
We now calculate the surface roughness exponent $\alpha$ by considering the following height-height correlation function \cite{PhdGiuraniuc} 
\begin{equation}
    \label{eq:height_correlation}
    \langle (h(x) - h(x+r))^2\rangle \propto r^{2\alpha}
\end{equation}
At distances $r>\lambda^{-1}$, since the deposition probabilities are independent, there is no possible dependence of the height difference on $r$ and $\langle\Delta h^2\rangle$ has a value which is given only by all the possible combinations: hole-hill, hole-nothing, hill-nothing. With increasing number of generations the number of combinations at each site increases; the height difference also increases but saturates as well to a value proportional to the total height (or depth). For shorter distances $r<\lambda^{-1}$ there are two possibilities for the height after the first generation: either they can be the same or differ by $\lambda^{-1}$ or $2\lambda^{-1}$, depending on whether the two points separated by $r$ are on the same block or not. The $r$-dependence of the height originates from the probability whether two points separated by a distance $r$ are on the same block or not, and this probability is proportional to $r$. Hence, $\langle \Delta h^2\rangle \sim r$. A similar reasoning can be made for $\lambda^{-2} < r < \lambda^{-1}$ and for all the other intervals. We can conclude that $\alpha = 1/2$ but that the slope changes at $\lambda^{-1},\, \lambda^{-2},\, \lambda^{-3}$ and so on. Hence, a fine structure emerges in the correlation, where the time evolution can be interpreted as successive magnifications, revealing more and more of this structure.

We have calculated the value of the roughness exponent $\alpha$ for $P = 0.2$, $Q = 0.1$ and $\lambda = 3$, which results in $\alpha = 0.485 \pm 0.0006$ and which is very close to the theoretically predicted value of $1/2$. It should be noted that this roughness exponent is a power law of the length from the beginning of the deposition process and consequently there is no initial growth that can be found as a power law of time. Hence, exponents $\beta$ or $z$ are nonexistent in our model.

These results are in stark contrast with the well-known models of random deposition and ballistic deposition where the exponents $\beta$ and $z$ do exist \cite{Edwards1982,KPZ,aharony2002}. In these deposition models it is assumed that all of the particles are of unit size and identical. While there have been studies on properties of surfaces resulting from the deposition of particles with varying size \cite{forgerini2009}, none consider the hyperbolic scaling and the simultaneous increase of the number of columns on the substrate. There are two fundamental differences between our model and these preexisting models. First, our model is \emph{synchronous}, meaning that all columns are visited simultaneously (i.e., in one generation), while other deposition models consider one particle being deposited at a time, updating time $t\rightarrow t+1$ when on average each column has been visited once. In our model, the probabilities $P$ and $Q$ control the number of deposition events in one generation. Second, in our hierarchical model, the number of available columns increases as $\lambda^n$ for increasing generations $n$. This signifies that we cannot define a uniform ``time" such as for the regular random deposition.

\section{Percolation in random hierarchical deposition on $D=1$}
\label{sec:percolation}       
Consider the hierarchical deposition model where now the hills are made of some conductive material such as copper or zinc, and an electric current is allowed to pass through the system from end to end. The substrate is assumed to be a perfect insulator. We now investigate whether a current is able to flow between the two endpoints and when a spanning cluster of conducting hills appears for the first time. This turns out to be connected to the  discussion of the previous section(s). We only consider percolation on a one-dimensional substrate, as the two-dimensional calculations involve a number of $2^n$ possible microscopic configurations and as such become very involved.

\begin{figure}[!htp]
    \centering
    \includegraphics[width=0.75\linewidth]{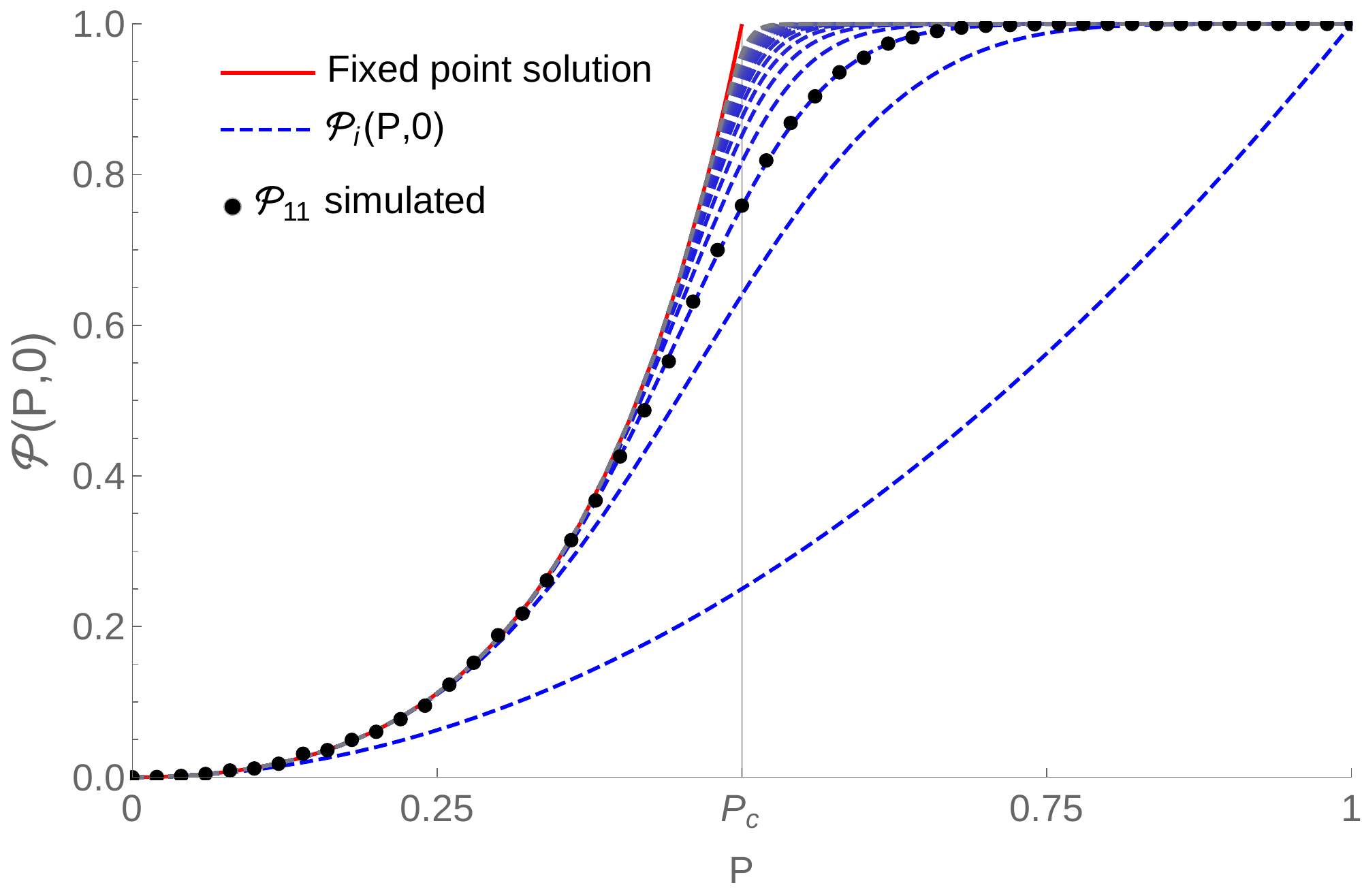}
    \caption{Percolation probability for $\lambda = 2$ and $Q=0$. The fixed-point solution is shown (red line) together with the percolation probabilities $\mathcal{P}_i$ for the first 100 generations in steps of five (blue, dashed lines) in ascending order, starting with $n=1$, and the simulated results for generation $n=11$ (black circles).}
    \label{fig:recurrence_2}
\end{figure}

\begin{figure}[!htp]
    \centering
    \includegraphics[width=0.75\linewidth]{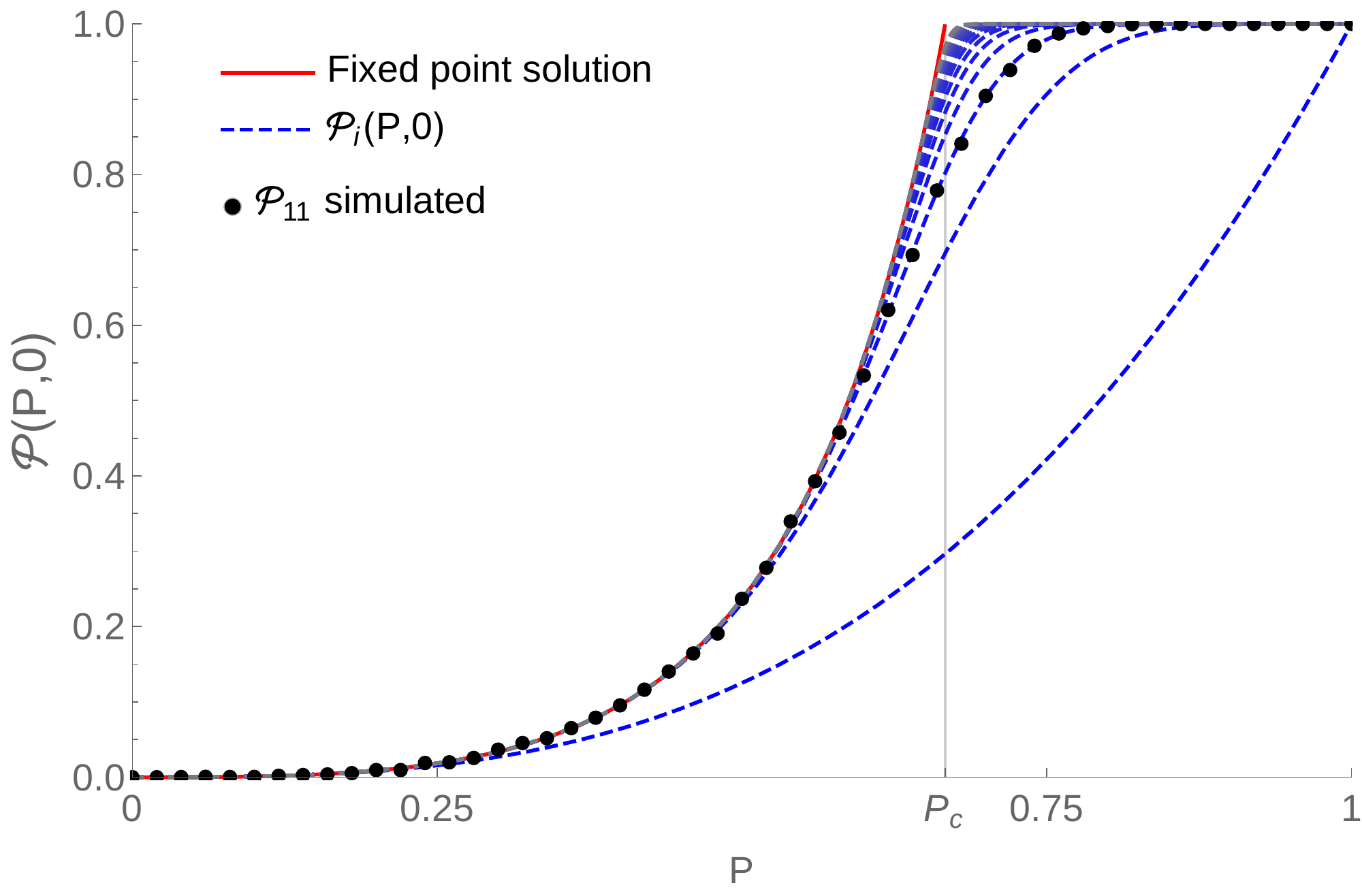}
    \caption{Percolation probability for $\lambda = 3$ and $Q=0$. The fixed-point solution is shown (red line) together with the percolation probabilities $\mathcal{P}_i$ for the first 100 generations in steps of five (blue, dashed lines) in ascending order, and the simulated results for generation $n=11$ (black circles).}
    \label{fig:recurrence_3}
\end{figure}

Percolation manifests itself when an uninterrupted chain of conducting hills is placed between the left and right sides of the unit interval $[0, 1]$. This can be realised in every generation $n>0$. Note that when in generation $n$ a hole is dug on a segment which did not experience any deposition up to order $n-1$, percolation is made impossible for every generation number exceeding $ n$. The probability to have reached percolation in generation $n$ or earlier is denoted by $\mathcal{P}_n(P,Q)$. We give here the example of $\lambda = 3$ but the results are valid for general $\lambda\geq2$.
\begin{itemize}
    \item Percolation in $n=0$ will be nonexistent. We assume the substrate is initially flat and has not experienced any deposition. Hence, by construction, $\mathcal{P}_0(P,Q)=0$.
    \item Percolation in $n = 1$. The only possibility for percolation already in the first step is when all $\lambda =3$ lattice sites are filled. So, $\mathcal{P}_1(P,Q) = P^3$.
    \item Percolation in $n=2$. Here are three possibilities: either 0, 1 or 2 hills have been deposited in the first generation, as shown in Fig.\ref{fig:test1}. The total percolation probability is the probability that percolation has already occurred in the previous generation added to the probability that percolation occurs in this generation, i.e.,
    \begin{equation}
        \mathcal{P}_2(P,Q) = P^3 + 3P^2 (1-S) P^3 + 3P (1-S)^2 P^6 + (1-S)^3 P^9\, .
    \end{equation}

    \begin{figure}[!ht]
    \centering
    \begin{subfigure}{.32\textwidth}
      \centering
      \includegraphics[width=.7\linewidth]{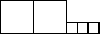}
      \caption{}
      \label{fig:sub1}
    \end{subfigure}%
    \begin{subfigure}{.32\textwidth}
      \centering
      \includegraphics[width=.7\linewidth]{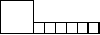}
      \caption{}
      \label{fig:sub2}
    \end{subfigure}
    \begin{subfigure}{.32\textwidth}
      \centering
      \includegraphics[width=.7\linewidth]{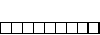}
      \caption{}
      \label{fig:sub3}
    \end{subfigure}
    \caption{(a) A possible configuration in which percolation is achieved when two blocks were deposited in $n=1$, with associated probability $3P^2 (1-S) P^3$. (b) Possible percolation when one block has been deposited, with probability $3P(1-S)^2 P^6$. (c) Possible percolation when no blocks have been deposited, with probability $(1-S)^3P^9$. }
    \label{fig:test1}
    \end{figure}
    \item Percolation in $n=3$. Some of the different possibilities are shown graphically in the appendix \ref{app:2}. The percolation probability is
    \begin{equation}
        \begin{split}
        \mathcal{P}_3(P,Q) &= \sum\limits_{k=0}^3{3 \choose k} P^{3-k} (1-S)^k \sum\limits_{i=0}^{3 k}{3 k \choose i} P^{3 k-i} (1-S)^i P^{3 i}\\
        &= \sum\limits_{k=0}^3 \sum\limits_{i=0}^{3 k} {3 \choose k} {3 k \choose i} P^{-(k+i)} (1-S)^{k+i} P^{3 (k+i)}\, .
        \end{split}
    \end{equation}
\end{itemize}
For general $\lambda$, the probability of percolation in $n=3$ is given by
\begin{equation}
    \begin{split}
    \mathcal{P}_3(P,Q) &= \sum\limits_{k=0}^\lambda{\lambda \choose k} P^{\lambda-k} (1-S)^k \sum\limits_{i=0}^{\lambda k}{\lambda k \choose i} P^{\lambda k-i} (1-S)^i P^{\lambda i}\\
    &= \sum\limits_{k=0}^\lambda \sum\limits_{i=0}^{\lambda k} {\lambda \choose k} {\lambda k \choose i} P^{-(k+i)} (1-S)^{k+i} P^{\lambda (k+i)}\, .
    \end{split}
\end{equation}
This can now be repeated for general $n$ and $\lambda$ to result in the following expression for the percolation probability
\begin{equation}
    \label{eq:nested_sums}
    \mathcal{P}_n(P,Q) = P^\lambda \sum\limits_{k_1=0}^\lambda \sum\limits_{k_2=0}^{\lambda k_1}\dots\sum\limits_{k_{n-1}=0}^{\lambda k_{n-2}} \, \prod\limits_{i=k_{n-1}}^{k_1}{\lambda k_{i-1} \choose i} \left[P^{\lambda-1} (1-S)\right]^{\sum\limits_{j=1}^{n-1}k_j}\, .
\end{equation}
The nested summations in \eqref{eq:nested_sums} indicate the strong memory effect pertaining to the history of deposition in  previous generations. Working out the above expressions explicitly starting with $\mathcal{P}_0 = 0$ gives 
\begin{equation}
    \label{eq:recurrence}
    \begin{split}
        \mathcal{P}_1(P,Q) &= P^\lambda\\
        \mathcal{P}_2(P,Q) &= \left(P+(1-S) P^\lambda\right)^\lambda\\
        \mathcal{P}_3(P,Q) &= \left(P+(1-S)\left(P+(1-S) P^\lambda\right)^\lambda\right)^\lambda\\
        &\vdots\\
        \mathcal{P}_n(P,Q) &= \left(P+(1-S)\mathcal{P}_{n-1}(P,Q)\right)^\lambda\, .
    \end{split}
\end{equation}
The fixed point of the recurrence equation \eqref{eq:recurrence} can be computed analytically for low values of $\lambda$ and numerically for larger values. In Fig. \ref{fig:recurrence_2} and \ref{fig:recurrence_3},  the first 100 iterations are shown for $\lambda = 2$ and $\lambda=3$ with $Q = 0$ together with numerically simulated results for $n=11$ averaged over 20000 realisations. Note that a singularity in the derivative develops at some values for $P$, which is reminiscent of a first-order phase transition.

The asymptotic behaviour for large $n$ of the recurrence relation \eqref{eq:recurrence} can be studied by considering a complementary problem for the hierarchical deposition model. Let us study the distribution of \textsl{empty} intervals in  generation $n$. We can identify the process of depositing a hill or an empty space with a rooted tree \cite{progress} where each vertex has $\lambda$ children and we select edges with probability $1-S$. The probability that a vertex has exactly $k$ children (i.e., empty disjoint subsets) is therefore
\begin{equation}
    \label{eq:prob_k_children}
    \xi_k = {\lambda \choose k} (1-S)^k P^{\lambda-k}\, ,
\end{equation}
so the expected number $\mu$ of empty subsets in the first generation is
\begin{equation}
    \label{eq:mean}
    \begin{split}
    \mu &= \sum\limits_{k=0}^\lambda k {\lambda \choose k} (1-S)^k P^{\lambda-k}\\
    &= \lambda (1-Q)^{\lambda-1}(1-S)\, .
    \end{split}
\end{equation}
We now define the stochastic variable $X_n$ as the number of vertices in generation $n$, with probability distribution $\mathbb{P}(X_n)$. The probability for a percolation cluster to form in generation $n$ is equal to the probability that the number of vertices for the branching process is equal to zero, i.e., there are no remaining empty subsets. Therefore, the connection between the original percolation problem and the complementary problem can be expressed as follows
\begin{equation}
    \mathcal{P}_n(P,Q) = \mathbb{P}(X_n =0)\, .
\end{equation}
To calculate the probability $\mathbb{P}(X_n =0)$, we define the following generating function $f$ for the sequence $\{\xi_k\}$,
\begin{equation}
    \label{eq:generating_function}
    \begin{split}
    f(x) &= \sum\limits_{k=0}^\lambda \xi_k x^k \\
    &= \sum\limits_{k=0}^\lambda x^k {\lambda \choose k} (1-S)^k P^{\lambda-k}\\
    &= (P+(1-S)x)^\lambda\, .
    \end{split}
\end{equation}
Note that this generating function has the same functional form as the recurrence relation \eqref{eq:recurrence}. Furthermore, it follows that $f(0) = P^\lambda$, $f(1) = (1-Q)^\lambda$ and $f'(1) = \mu$. Now, define the conditional probabilities $\xi_j^{(n)} = \mathbb{P}(X_{n} = j|X_0=1)$. The generating function for this new process is
\begin{equation}
    \label{eq:generating_function_composed}
    F_n = f^{(n)} \equiv f\circ f\circ...f\, ,
\end{equation}
which is the $n$-fold composition of the generating function $f$ with itself. Therefore, assuming that $X_0 = 1$, the probability $\mathbb{P}(X_n=0)$ is the following function
\begin{equation}
    \label{eq:prob_zero}
    \mathbb{P}(X_n=0) = f^{(n)}(0)\, .
\end{equation}
This implies that for $n\rightarrow\infty$ the iteration converges to the first fixed point of $f(x)$ that is reached when starting from $x=0$. If $\mu\leq1$, the sequence converges to 1, indicating percolation while for $\mu>1$ the iteration converges to another fixed point. Hence, the percolation threshold $P_c$ can be calculated as follows
\begin{equation}
    \label{eq:percolation_treshold_full}
        P_c = 1-Q-\frac{(1-Q)^{1-\lambda}}{\lambda}\, ,
\end{equation}
and for $Q=0$ this reduces to
\begin{equation}
\label{eq:percolation_treshold}
    P_c = 1-\frac{1}{\lambda}\, .
\end{equation}
For $\lambda=2$ and $\lambda=3$, the percolation thresholds are, respectively, $P_c = 1/2$ and $P_c = 2/3$, as can be seen in Fig. \ref{fig:recurrence_2} and \ref{fig:recurrence_3}. Note that the above threshold value \eqref{eq:percolation_treshold} is the same value as was found for $S$ in the previous section, for the separation point between the Euclidean and the fractal regimes for the number of coastal points. Since now $Q=0$, $S=P=P_c$ is precisely the condition for logarithmic fractality. 

\begin{figure}[ht]
    \centering
    \includegraphics[width = 0.75\linewidth]{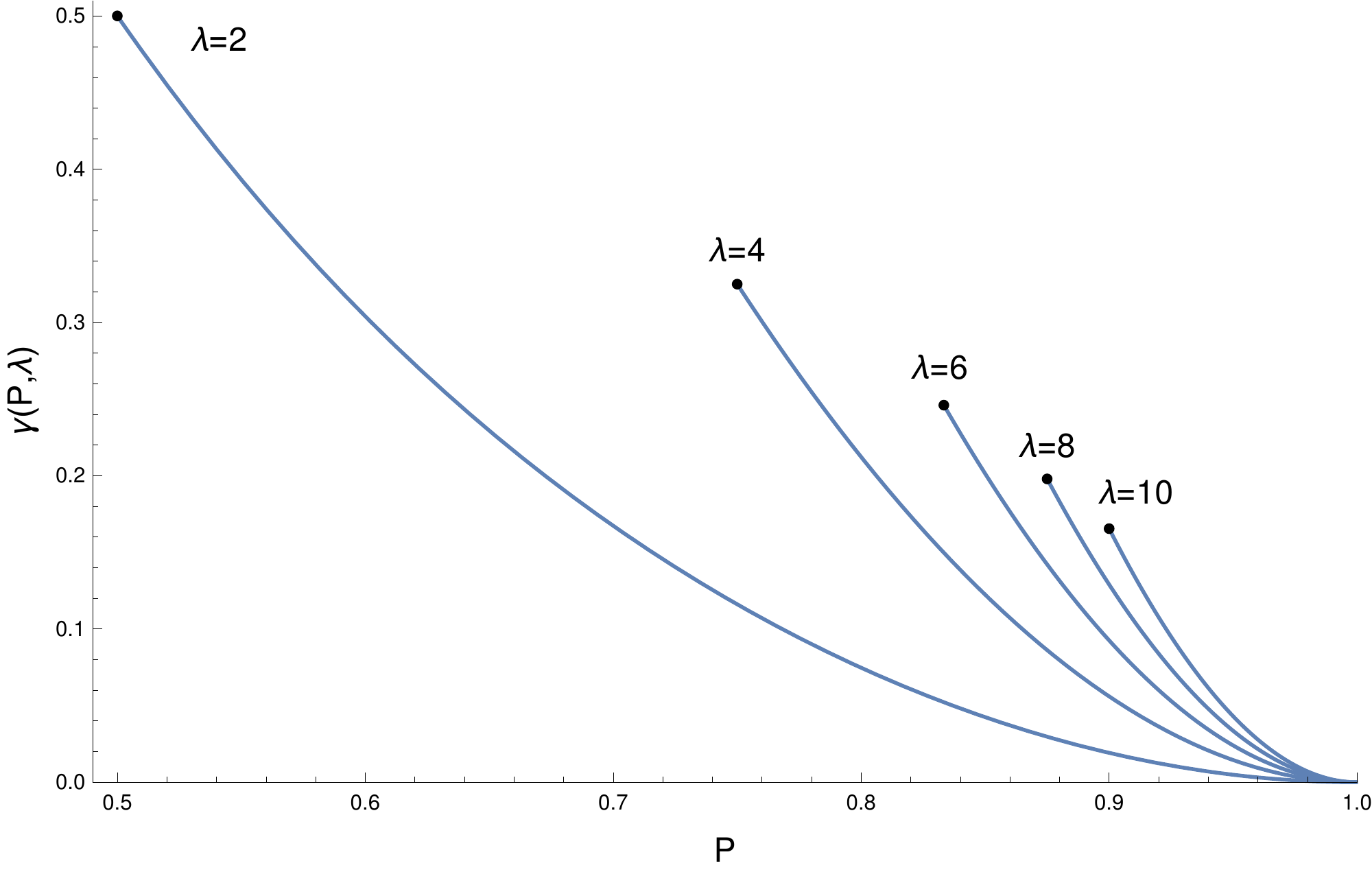}
    \caption{The constant $\gamma$ describing the asymptotic average of the number of coastal points for $P>P_c$ for $\lambda\in\{2,4,6,8,10\}$. The black dots indicate $P=P_c$ for each value of $\lambda$.}
    \label{fig:gammaplot}
\end{figure}

This coincidence can be understood from the following correspondence. When percolation occurs, the sea level is covered with conducting blocks, which, on average, inhibit the formation of new coastal points and, at the same time, destroy existing coastal points. For $P>P_c$ and $n\rightarrow\infty$, percolation is almost certainly achieved and the number of coastal points enters the Euclidean regime where it saturates on average to a constant value $\gamma <1$, as shown in Section \ref{sec:coastal} and Appendix \ref{app:1}. This $\gamma$ is the following number (for $Q=0$):
\begin{equation}
    \gamma =  \frac{2\lambda (\lambda-1) (1-P)^2 (1-P+P^2)}{\lambda-1+P}
\end{equation}
For $P\leq P_c$, the number of coastal points grows either exponentially, for  $P < P_c$, or linearly, for $P = P_c$. However, for $P>P_c$ this number effectively becomes zero, as can be seen in Fig. \ref{fig:gammaplot} for different values of $\lambda$. In this regime, the number of coastal points vanishes, since its asymptotic average is $\gamma\ll1$ (in a statistical sense) and percolation is almost certainly achieved. This explains the precise correspondence between coastal-point non-proliferation and percolation.

The calculation involving the rooted tree of empty subsets can be directly mapped to the study of the random Cantor set \cite{Falconer1994OnTG, Orzechowski1997}, and which we believe to be possible to extend to higher dimensions with moderate effort. The above results are valid for $Q=0$. When $Q>0$, the probability for a percolation cluster to form is never equal to one so a percolation threshold is nonexistent. Therefore, a singularity is absent for the fixed point of the recurrence relation \eqref{eq:recurrence}.

The functional form of the solution of equation \eqref{eq:recurrence} for $P\leq P_c$ can be directly calculated by finding the fixed-point solution $\mathcal{P}_\infty$, which is possible for low values of $\lambda$. Note that $\mathcal{P}_\infty = 1$ is the trivial fixed-point solution for $P\geq P_c$. We will denote the nontrivial solution by $\theta_\lambda(P)$ and we assume $Q=0$. Hence, for $\lambda=2$, the nontrivial solution $\theta_2(P)$ of
\begin{equation}
    \mathcal{P}_\infty = \left(P + (1-P) \mathcal{P}_\infty\right)^2
\end{equation}
is
\begin{equation}
    \label{eq:FP_L2}
    \theta_2(P) = \left(\frac{P}{1-P}\right)^2\, ,
\end{equation}
while for $\lambda=3$, the nontrivial solution $\theta_3(P)$ is
\begin{equation}
    \label{eq:FP_L3}
    \theta_3(P) = 1-\frac{3}{2 (1-P)} + \frac{1}{2}\sqrt{\frac{1+3P}{(1-P)^3}}\, .
\end{equation}
We now study the critical behaviour at $P_c$, which we expect to have the asymptotic form
\begin{equation}
    \label{eq:general_form}
    1 - \theta_\lambda(P)  \sim c(P_c-P)^\beta \, , \qquad P\uparrow P_c 
\end{equation}
for some $c>0$ and $\beta>0$. The critical behaviour of the solution at the fixed point can be found by expanding $\theta_\lambda$ about the percolation threshold, i.e.,
\begin{equation}
    \label{eq:expansion_FP}
    \begin{split}
        \theta_2(P) &= 1 + 8 (P-P_c) +\mathcal{O}((P-P_c)^2)\\
        \theta_3(P) &= 1 + 9 (P-P_c) + \mathcal{O}((P-P_c)^2)\\
        \theta_4(P) &= 1 + \frac{32}{3}(P-P_c) + \mathcal{O}((P-P_c)^2)\\
        &\vdots\\
        \theta_\lambda(P) &= 1+ \frac{2\lambda^2}{\lambda-1}(P-P_c) +\mathcal{O}((P-P_c)^2)\, .
    \end{split}
\end{equation}
Hence, the critical exponent $\beta = 1$ is obtained. This singularity is reminiscent of a first-order phase transition in view of the jump  in the first derivative of $\mathcal{P}_\infty(P)$ at $P_c$.

\section{Percolation with alternating deposition probabilities}\label{sec:percolation_alternating_prob}      
It is possible for the hills (and holes) to originate from different sources, thereby changing the resulting landscape. First, we will study one such system where two sources of deposition are present and for which the characteristic length scales remain the same, i.e., $\lambda_1=\lambda_2 =\lambda$. This has previously been studied in the context of evolving landscapes in the periodical extension of the hierarchical deposition model and logarithmic fractal geometry was confirmed  \cite{POSAZHENNIKOVA2000}. As an extension of this previous study, we now explore percolation properties. We consider two sources with deposition probabilities $(P_1,Q_1)$ and $(P_2,Q_2)$.

\begin{figure}[!htp]
    \centering
    \includegraphics[width=0.7\linewidth]{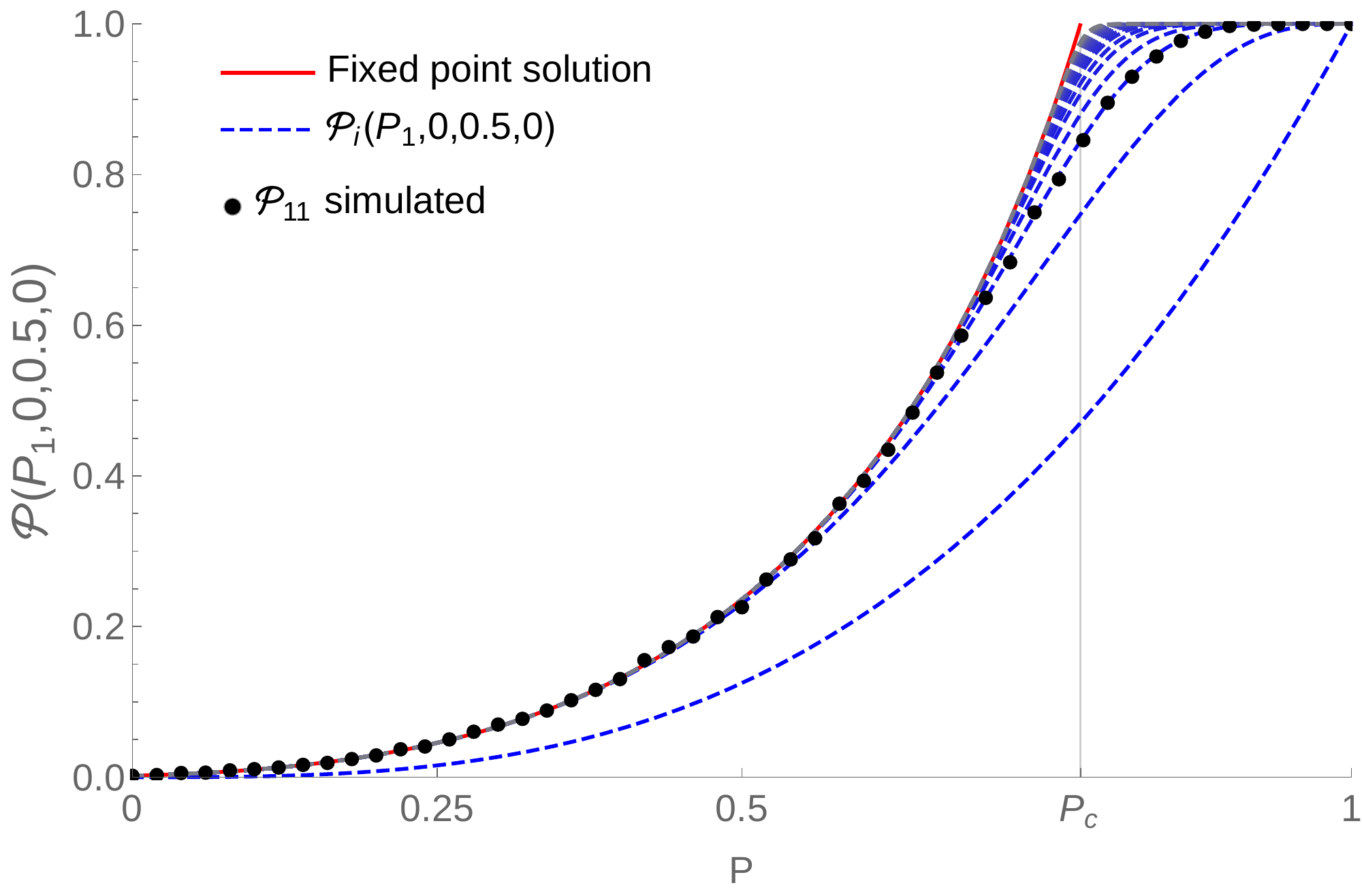}
    \caption{Percolation probability $\mathcal{P}_n(P_1,Q_1,P_2,Q_2)$ for alternating deposition probabilities with $\lambda = 3$, $Q_1= Q_2=0$ and $P_2 = 1/2$, as a function of $P \equiv P_1$. The fixed-point solution is shown (red line) together with the percolation probabilities for the first 100 generations in steps of five (blue, dashed lines) in ascending order, and the simulated results for $n=11$ (black circles).}
    \label{fig:perc_alt_prob}
\end{figure}

Repeating the calculations from the previous section \ref{sec:percolation}, now with alternating probabilities, it is straightforward to see that the percolation probability $\mathcal{P}_n(P_1,Q_1,P_2,Q_2)$ for the first four generations is given by, with $S_i = P_i + Q_i$,
\begin{equation}
    \label{eq:alternating_percolation_examples}
    \begin{split}
        \mathcal{P}_1(P_1,Q_1,P_2,Q_2) &= P_1^\lambda\\
        \mathcal{P}_2(P_1,Q_1,P_2,Q_2) &= \left[P_1 + P_2^\lambda\left(1-S_1\right)\right]^\lambda\\
        \mathcal{P}_3(P_1,Q_1,P_2,Q_2) &= \left[P_1 +\left(1-S_1\right)\left(P_2 + P_1^\lambda(1-S_2)\right)^\lambda\right]^\lambda\\
        \mathcal{P}_4(P_1,Q_1,P_2,Q_2) &= \left[P_1 +\left(1-S_1\right)\left(P_2 + (1-S_2)\left(P_1 + P_2^\lambda (1-S_1)\right)^\lambda\right)^\lambda\right]^\lambda\, .
    \end{split}
\end{equation}
Generalising this procedure, the percolation probability for generation $n$ is then
\begin{equation}
    \label{eq:alternating_percolation_general}
    \begin{split}
        \mathcal{P}_n(P_1,Q_1,P_2,Q_2) &=\left[P_1 + (1-S_1) \left(P_2 +(1-S_2) \mathcal{P}_{n-2}\right)^\lambda\right]^\lambda\, .
    \end{split}
\end{equation}
From this expression one can see that when $P_1=P_2=P$ and $Q_1=Q_2=Q$ the percolation probability reduces to that of the uniform hierarchical random deposition model \eqref{eq:recurrence}. Continuing as in the previous section, the expected number of empty subsets can be calculated in a similar manner, i.e.,
\begin{equation}
    \label{eq:mean_alternating_probability}
    \begin{split}
        \mu &= \sum\limits_{k=0}^\lambda {\lambda \choose k} P_1^{\lambda-k} (1-S_1)^k \sum\limits_{i=0}^{\lambda k} i {\lambda k \choose i} P_2^{\lambda k-i} (1-S_2)^i\\
        &= \lambda^2 (1-S_1)(1-S_2)\left((1-Q_2)\left(P_1 + (1-S_1)(1-Q_2)^\lambda\right)\right)^{\lambda-1}\, .
    \end{split}
\end{equation}
From equation \eqref{eq:mean_alternating_probability} the percolation threshold can once again be calculated for either $P_1$ or $P_2$. With $Q_1 = Q_2 = 0$, the percolation threshold becomes
\begin{equation}
    \label{eq:percolation_threshold_alternating}
    P_{1,c} = 1-\frac{1}{\lambda^2 (1-P_2)}\, .
\end{equation}
For $P_2$ a multiple of $P_1$, i.e., for $P_2 = r P_1$, $r\in\mathbb{R}^+$, the percolation threshold is
\begin{equation}
    P_{1,c} = \frac{r+1}{2r}-\sqrt{\frac{1}{r\lambda^2} + \left(\frac{r-1}{2r}\right)^2}\, .
\end{equation}
Notice that this expression reduces to the percolation threshold \eqref{eq:percolation_treshold} for the uniform random deposition model when $r=1$. In Fig. \ref{fig:perc_alt_prob}, the percolation probability is shown for the first 100 iterations of the alternating model with $\lambda = 3$ together with the fixed-point solution. The percolation threshold is $P_{1,c} = 7/9$ and the numerically simulated results for $n=11$ are averaged over 20000 realisations. The probability to dig a hole is assumed to be zero for both sources, i.e., $Q_1 = Q_2 = 0$, while the probability to deposit a hill in the second generation is $P_2 = 1/2$.

\section{Percolation with alternating rescaling factors}\label{sec:alternating_scale}
We now assume the probabilities $(P,Q)$ to be constant but take the characteristic length rescaling factors $\lambda_1$ and $\lambda_2$ to be different, as was initially proposed in \cite{Indekeu2001}. The percolation probability for the first three generations can be calculated in the same manner as before, resulting in
\begin{equation}
    \label{eq:alternating_percolation_scale_examples}
    \begin{split}
        \mathcal{P}_1(P,Q,\lambda_1,\lambda_2) &= P^{\lambda_1}\\
        \mathcal{P}_2(P,Q,\lambda_1,\lambda_2) &= \left[P + P^{\lambda_2}\left(1-S\right)\right]^{\lambda_1}\\
        \mathcal{P}_3(P,Q,\lambda_1,\lambda_2) &= \left[P +\left(1-S\right)\left(P + P^{\lambda_1}(1-S)\right)^{\lambda_2}\right]^{\lambda_1}\, .
    \end{split}
\end{equation}

The percolation probability for generation $n$ can be found by continuing the above sequence
\begin{equation}
    \label{eq:alternating_percolation_scale_general}
    \begin{split}
        \mathcal{P}_n(P,Q,\lambda_1,\lambda_2) &=\left[P + (1-S) \left(P +(1-S) \mathcal{P}_{n-2}\right)^{\lambda_2}\right]^{\lambda_1}\, ,
    \end{split}
\end{equation}
which reduces to the uniform random deposition model for $\lambda_1 = \lambda_2 = \lambda$. The average number of empty subsets after one generation, $\mu$,  is given by 
\begin{equation}
    \label{eq:mean_alternating_scale}
    \begin{split}
        \mu &= \sum\limits_{k=0}^{\lambda_1} {\lambda_1 \choose k} P^{\lambda_1 -k} (1-S)^k \sum\limits_{i=0}^{\lambda_2 k} i {\lambda_2 k \choose i} P^{\lambda_2 k-i} (1-S)^i\\
        &= \lambda_1 \lambda_2 (1-S)^2 (1-Q)^{\lambda_2-1}\left(P + (1-S)(1-Q)^{\lambda_2}\right)^{\lambda_1-1}\, .
    \end{split}
\end{equation}
The percolation threshold $P_c$ can be calculated by solving $\mu =1$ for $P$, i.e., 
\begin{equation}
    \label{eq:percolation_threshold_alternating_scale}
    P_c = 1-\frac{1}{\sqrt{\lambda_1 \lambda_2}}\, ,
\end{equation}
which reduces to the formerly calculated expression \eqref{eq:percolation_treshold} in the uniform random deposition model in which the rescaling factors are equal, $\lambda_1=\lambda_2=\lambda$. Note that for the calculation of the percolation threshold the roles of $\lambda_1$ and $\lambda_2$ in \eqref{eq:mean_alternating_scale} can be freely interchanged. 

\begin{figure}[!htp]
    \centering
    \includegraphics[width=0.7\linewidth]{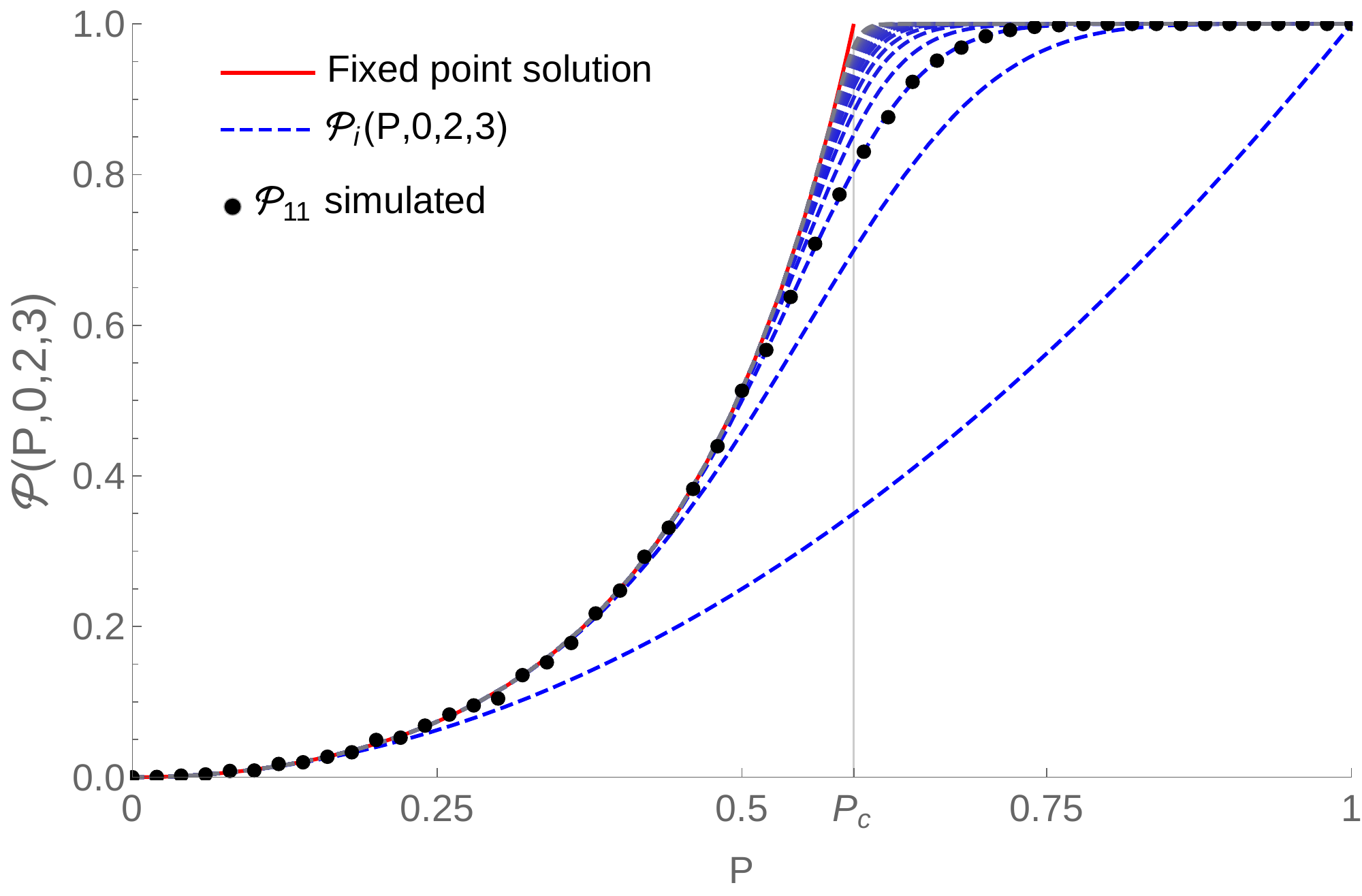}
    \caption{Percolation probability $ \mathcal{P}_n(P,Q,\lambda_1,\lambda_2)$ for alternating rescaling factor deposition with $\lambda_1 = 2$, $\lambda_2 = 3$, $Q=0$. The fixed-point solution is shown (red line) together with the percolation probabilities for the first 100 generations in steps of five (blue, dashed lines) in ascending order, and the simulated results for $n=11$ (black circles).}
    \label{fig:perc_alt_scale}
\end{figure}

In Fig.\ref{fig:perc_alt_scale}, the percolation probability is shown for $\lambda_1 = 2$ and $\lambda_2 = 3$ together with the fixed-point solution of the corresponding recurrence relation \eqref{eq:alternating_percolation_scale_general} and numerically simulated results for $n=11$ averaged over 20000 realisations.. The percolation threshold $P_c = 1-1/\sqrt{6}$ is also shown.

\section{Conclusions}

In this paper we have studied the problem of the enumeration of coastal points (or coastlines) and the problem of percolation in a model for random hierarchical deposition. We have pointed out that the two problems are closely connected. A first key result is the occurrence of a non-universality in that three distinct topological regimes are found. Either the number of coastal points formed saturates to a constant value with increasing generation number $n$, indicating Euclidean behaviour, or it increases exponentially, indicating a fractal regime with fractal dimension $D_f$. The intermediate point between these two regimes is characterised  by a linear increase in the number of coastal points, which marks a logarithmic fractal behaviour. This behavior, which constitutes a special case here, was, in contrast, previously found to be a universal property of the geometry of the landscape produced by a hierarchical deposition process. We have found that the intermediate point is characterized by a simple relationship between the deposition probabilities and the length rescaling factor of the model. The geometrical properties of the model thus have turned out to be surprisingly rich. We have briefly studied the surface morphology and calculated a roughness exponent $\alpha$, which we confirmed numerically. Future research could explore the dynamical scaling of the surface and investigate possible connections with related interface growth models \cite{Family_1985,KPZ}. 

In the second part of the paper we have studied the percolation properties of hierarchical deposition of conducting blocks and shown that the asymptotic percolation threshold probability $P_c$ is located at exactly the same intermediate-point value for the deposition probability that separates Euclidean from fractal behavior. We have interpreted and understood this close connection by relating the vanishing of the number of coastal points in the Euclidean regime to the percolated phase of the conducting deposit. Conversely, the fractal regime, in which the  coastal points are prolific, is essentially non-percolating. 

Finally, we extended the percolation calculations to cases in which either the deposition probabilities or the rescaling factor alternate (periodically) between two values. Further research could include extending the coastal-point calculations to the modified deposition models of sections \ref{sec:percolation_alternating_prob} and \ref{sec:alternating_scale} or for a magnetic version of the model which has been studied in \cite{magneticDeposition}. In a more applied arena, optical and electromagnetic properties of the surface can be studied and tested in real-world applications such as the design of antennae \cite{Kakkar2019} or acoustic metamaterials \cite{MAN2019}. For these applications, it could be worthwhile to investigate the properties of the hierarchical random deposition model in different geometries, e.g., triangular, spherical, hexagonal etc., and with different boundary conditions.

\section{Dedication}

We dedicate this paper to the memory of Dietrich Stauffer, who was a grandmaster in computational statistical physics. One of us (J.O.I.) is especially grateful for professor Stauffer's frank and encouraging comments by virtue of which  a simple and modest idea could be developed into something worthwhile and significant. 

\newpage 
\appendix
\section{Number of coastal points}\label{app:1}
Starting from equation \eqref{eq:coastal_points} and working out the sums explicitly results in the following expression for the number of coastal points in generation $n$:
\begin{equation}
    \label{eq:coastal_points_worked}
    \begin{split}
        \mathcal{N}_n(P,Q) &= 2 \lambda P(1-P)\frac{1-(1-S)^{2n}}{1-(1-S)}\\
        &+ 2\lambda(\lambda-1)P(1-P)(1-S) \left[\frac{1-\left(\lambda (1-S)\right)^{n-1}}{1-\lambda(1-S)}\right]\\
        &+ 2\lambda P(Q-P) (1-S) \left[\frac{1-\lambda^{n-1} (1-S)^{n-1}}{1-\lambda(1-S)}\right]\left[\frac{1-\lambda^{-(n-1)} (1-S)^{n-1}}{1-\lambda^{-1}(1-S)}\right]\\
        &+ 2\lambda(\lambda-1) P(1-P)(1-S)^3 \left[\frac{1-\lambda^{n-2} (1-S)^{n-2}}{1-\lambda(1-S)}\right]\left[\frac{1-\lambda^{-(n-2)} (1-S)^{n-2}}{1-\lambda^{-1}(1-S)}\right]
    \end{split}
\end{equation}

It can be seen now that if $\lambda (1-S)<1$, the above expression converges to a constant $\gamma$ for very large values of $n$. This constant can be calculated to be
\begin{equation}
    \label{eq:gamma}
    \begin{split}
        \gamma &= \frac{2 \lambda P(1-P)}{1-(1-S)}\\
        &+ \frac{2\lambda(\lambda-1)P(1-P)(1-S)}{1-\lambda(1-S)}\\
        &+ \frac{2\lambda P(Q-P) (1-S) }{1-(1-S)(\lambda+\lambda^{-1}) +(1-S)^2}\\
        &+ \frac{2\lambda(\lambda-1)P(1-P) (1-S)^3 }{1-(1-S)(\lambda+\lambda^{-1}) +(1-S)^2}\\
    \end{split}
\end{equation}

For $\lambda (1-S)>1$, the number of coastal points increases exponentially. The limiting case $\lambda (1-S)=1$ results in a linear increase in coastal points. This can be seen by formally calculating the limit of the number of coastal points \eqref{eq:coastal_points_worked} for $\lambda(1-S)\rightarrow1$, which results in
\begin{equation}
    \begin{split}
        \mathcal{N}_n(P,Q) &= 2 \lambda (\lambda-1)P(1-P)(1-S)(n-1)\\
        &+ 2\lambda P(Q-P)(1-S) (n-1)\\
        &+ 2\lambda(\lambda-1)P(1-P)(1-S)^3 (n-2)
    \end{split}
\end{equation}

\section{Percolation in third generation}\label{app:2}
In generation $n=3$ with $\lambda=3$ there are 721 different possibilities to obtain percolation. We will not list them all but will show some possibilities and comment on the degeneracy within one combination. 

\begin{figure}[ht]
    \centering
    \begin{subfigure}{.32\textwidth}
      \centering
      \includegraphics[width=.7\linewidth]{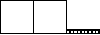}
      %\caption{$3P^2 (1-S)\cdot (1-S)^3\cdot P^9$}
      \caption{}
      \label{fig:sub4}
    \end{subfigure}%
    \begin{subfigure}{.32\textwidth}
      \centering
      \includegraphics[width=.7\linewidth]{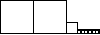}
      %\caption{$3P^2 (1-S)\cdot 3P(1-S)^2\cdot P^6$}
      \caption{}
      \label{fig:sub5}
    \end{subfigure}
    \begin{subfigure}{.32\textwidth}
      \centering
      \includegraphics[width=.7\linewidth]{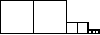}
      %\caption{$3P^2 (1-S)\cdot 3P^2(1-S)\cdot P^3$}
      \caption{}
      \label{fig:sub6}
    \end{subfigure}
    \caption{Possible configurations in which percolation is achieved when two blocks were deposited in $n=1$ and (a) no block was deposited, (b) one block was deposited or (c) two blocks were deposited in $n=2$.}
    \label{fig:test2}
\end{figure}

First, consider combinations that experienced the deposition of two hills of length $1/\lambda$ and one empty space in the first generation. There are only three ways to do this. In the second generation, consider hills of length $1/\lambda^2$ or empty spaces being deposited in the empty space created in the first generation. Once again, only zero, one or two hills can be deposited, with the possible number of configurations being respectively 1, 3 and 3. In the third generation, percolation can only occur when all remaining empty spaces at sea level are filled with conducting hills. These possibilities are shown in Fig.\ref{fig:test2}. It is now straightforward to see that this results in 21 different possible configurations. Added together, these probabilities result in
\begin{equation}
    3 P^2 (1-S) \sum\limits_{k=0}^2{2\choose k}P^k (1-S)^{3-k}P^{3 (3-k)}
\end{equation}

\begin{figure}[ht]
    \centering
    \begin{subfigure}{.32\textwidth}
      \centering
      \includegraphics[width=.7\linewidth]{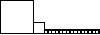}
      \caption{$3P (1-S)^2\cdot6P (1-S)^5\cdot P^{15}$}
      \label{fig:sub7}
    \end{subfigure}%
    \begin{subfigure}{.32\textwidth}
      \centering
      \includegraphics[width=.7\linewidth]{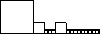}
      \caption{$3P(1-S)^2\cdot 15P^2(1-S)^4\cdot P^{12}$}
      \label{fig:sub8}
    \end{subfigure}
    \begin{subfigure}{.32\textwidth}
      \centering
      \includegraphics[width=.7\linewidth]{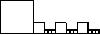}
      \caption{$3P(1-S)^2\cdot 20P^3(1-S)^3\cdot P^{9}$}
      \label{fig:sub9}
    \end{subfigure}
    \caption{Possible configurations in which percolation is achieved when one block was deposited in $n=1$ and (a) one block was deposited, (b) two blocks were deposited or (c) three blocks were deposited in $n=2$. The possibilities where 0, 4 or 5 blocks were deposited are not shown.}
    \label{fig:test3}
\end{figure}

Next, we consider possibilities in which a single hill was deposited in the first generation. Once again there are only three possibilities. In the second generation, 0 to 5 hills can be deposited with the number of configurations being, respectively, 1, 6, 15, 20, 15 and 6. In the third generation, empty spaces at sea level need to be filled with conducting hills to obtain percolation. Some combinations are shown in Fig.\ref{fig:test3} for 1, 2 or 3 hills being deposited in the second generation.

This results in a total of 189 different microscopic configurations. Once again adding these probabilities together, the following expression is obtained
\begin{equation}
    3P(1-S)^2 \sum\limits_{k=0}^5{5\choose k} P^k (1-S)^{6-k}P^{18-3k}
\end{equation}

Lastly, consider the possibilities in which in the first generation nothing has been deposited. We will not show this here but after some calculations it is straightforward to see that this results in 511 unique combinations for percolation. In this case the probability becomes
\begin{equation}
    (1-S)^3 \sum\limits_{k=0}^8{8\choose k} P^k (1-S)^{9-k}P^{27-3k}
\end{equation}

In total, for $n=3$, there are 721 different microscopic combinations possible to obtain percolation, with an associated probability
\begin{equation}
    \label{eq:third_gen}
\begin{split}
    3 P^2 (1-S) \sum\limits_{k=0}^2 & {2\choose k}P^k (1-S)^{3-k}P^{9-3k}\\
    + 3P(1-S)^2 \sum\limits_{k=0}^5 & {5\choose k} P^k (1-S)^{6-k}P^{18-3k}\\ 
    + (1-S)^3 \sum\limits_{k=0}^8 & {8\choose k} P^k (1-S)^{9-k}P^{27-3k}
\end{split}
\end{equation}
Adding the different contributions results in an expression for the percolation probability for $\lambda = 3$ and $n=3$.

\bibliography{percolation.bib}
\end{document}